\documentclass[twocolumn]{aastex63}
\usepackage{amsmath}
\usepackage{xcolor}

\def \kms {km\,s$^{-1}$}
\graphicspath{{./}{figures/}}
\usepackage{xcolor, soul}
\sethlcolor{green}
\usepackage{graphicx}

\begin{document}

\title{The Center of Expansion and Age of the Oxygen-rich Supernova Remnant 1E 0102.2-7219}

\correspondingauthor{John Banovetz}
\email{jbanovet@purdue.edu}

\author[0000-0003-0776-8859]{John Banovetz}
\affil{Department of Physics and Astronomy, Purdue University, 525 Northwestern Avenue, West Lafayette, IN 47907, USA}

\author[0000-0002-0763-3885]{Dan Milisavljevic}
\affil{Department of Physics and Astronomy, Purdue University, 525 Northwestern Avenue, West Lafayette, IN 47907, USA}

\author{Niharika Sravan}
\affil{Department of Physics and Astronomy, Purdue University, 525 Northwestern Avenue, West Lafayette, IN 47907, USA}

\author[0000-0003-3829-2056]{Robert A.\ Fesen}
\affil{Department of Physics and Astronomy, 6127 Wilder Laboratory, Dartmouth College, Hanover, NH 03755, USA}

\author[0000-0002-7507-8115]{Daniel J.\ Patnaude}
\affil{Center for Astrophysics \textbar\  Harvard \& Smithsonian, 60 Garden Street, Cambridge, MA 02138, USA}

\author[0000-0002-7507-8115]{Paul P.\ Plucinsky}
\affil{Center for Astrophysics \textbar\ Harvard \& Smithsonian, 60 Garden Street, Cambridge, MA 02138, USA}

\author[0000-0003-2379-6518]{William P.\ Blair}
\affil{Department of Physics \& Astronomy, Johns Hopkins University, Baltimore, MD 21218, USA}

\author[00000-0002-4471-9960]{Kathryn E.\ Weil}
\affil{Department of Physics and Astronomy, Purdue University, 525 Northwestern Avenue, West Lafayette, IN 47907, USA}

\author{Jon A.\ Morse}
\affil{BoldlyGo Institute, 1370 Broadway 5th Floor Suite 572, New York, NY 10018, USA}

\author[0000-0003-4768-7586]{Raffaella Margutti}
\affil{Center for Interdisciplinary Exploration and Research in Astrophysics (CIERA) and Department of Physics and Astronomy, Northwestern University, Evanston, IL 60208, USA)}

\author[0000-0001-7081-0082]{Maria R.\ Drout}
\affiliation{David A. Dunlap Department of Astronomy and Astrophysics, University of Toronto, 50 St.\ George Street, Toronto, Ontario, M5S 3H4, Canada}
\affiliation{Observatories of the Carnegie Institute for Science, 813 Santa Barbara Street, Pasadena, CA 91101-1232, USA}

\begin{abstract}

We present new proper motion measurements of optically emitting oxygen-rich knots of supernova remnant 1E 0102.2-7219 (E0102), which are used to estimate the remnant's center of expansion and age. Four epochs of high resolution {\sl Hubble Space Telescope} images spanning 19 yr were retrieved and analyzed. We found a robust center of expansion of $\alpha$=1$^{h}$04$^{m}$02.48$^{s}$ and $\delta$=-72$^{\circ}$01$^{\prime}$53.92$^{\prime\prime}$ (J2000) with 1-$\sigma$ uncertainty of 1.77$^{\prime\prime}$ using 45 knots from images obtained with the Advanced Camera for Surveys using the F475W filter in 2003 and 2013 having the highest signal-to-noise ratio. 
We also estimate an upper limit explosion age of 1738 $\pm$ 175 yr by selecting knots with the highest proper motions, that are assumed to be the least decelerated. We find evidence of an asymmetry in the proper motions of the knots as a function of position angle. We conclude that these asymmetries were most likely caused by interaction between E0102's original supernova blast wave and an inhomogeneous surrounding environment, as opposed to intrinsic explosion asymmetry. The observed non-homologous expansion suggests that the use of a free expansion model inaccurately offsets the center of expansion and leads to an overestimated explosion age. We discuss our findings as they compare to previous age and center of expansion estimates of E0102 and their relevance to a recently identified candidate central compact object.

\end{abstract}

\keywords{ISM: individual(SNR 1E 0102.2-7219)--
ISM: kinematics and dynamics -- supernova remnants}

\section{Introduction} \label{sec:intro}

\begin{figure*}[!htp]
\centering
\includegraphics[width=0.8\textwidth, angle=0]{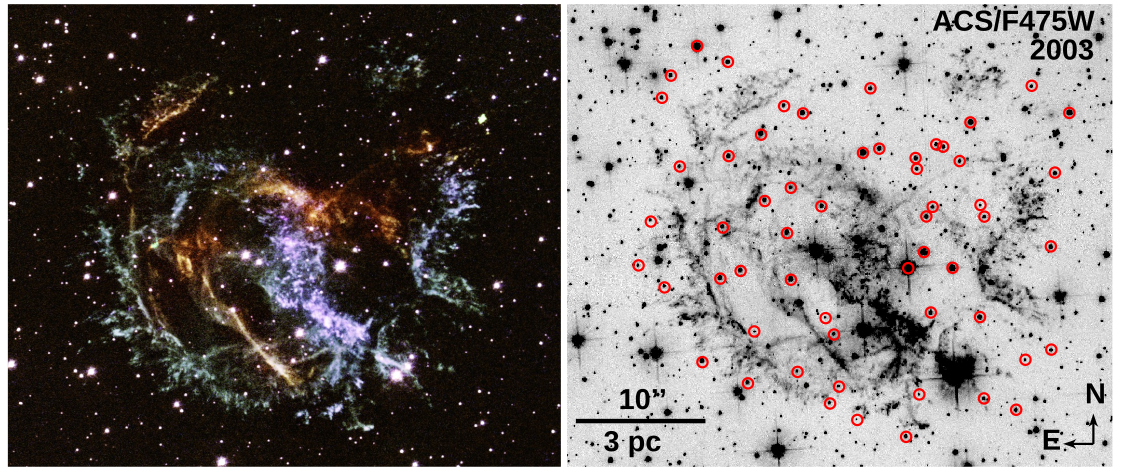}
\caption{Left: Composite image of E0102 made from the FQ492N (blue), F502N (green), and FQ508N (red) filters obtained in 2014 with WFC3/UVIS. The filters are sensitive to [O~III] 5007 emission with Doppler velocities less than $-2000$ \kms, between $-2000$ and $+2500$ \kms, and greater than $+2500$ \kms, respectively. Right: 2003 ACS/F475W image of E0102 sensitive to all velocities of [O~III] $\lambda\lambda$4959, 5007 emission. Red circles mark reference stars used to align all epochs of images.}
\label{fig:geomapref}
\end{figure*}

Supernova remnants (SNRs) encode valuable information about the explosion processes of supernovae and their progenitor systems (see \citealt{Milisavljevic2017} for a review). Young ($\lessapprox 2000$ yr), nearby ($< 1$ Mpc) oxygen-rich (O-rich) SNRs, created from the collapsed cores of massive stars ($> 8M_{\odot}$; \citealt{Smartt2009}), are particularly well-suited laboratories to study details of supernova explosion dynamics, as the kinematic and chemical properties of the metal rich debris can retain details of the parent supernova explosion \citep{Blair2000,Flanagan2004}. Ejecta can be followed over many years to determine the precise origin of the explosion, which in turn can be used to estimate the age of the remnant since explosion. Furthermore, interaction between the supernova's blast wave and ejecta with surrounding circumstellar and interstellar material (CSM/ISM) can constrain mass loss and evolutionary transitions experienced by the progenitor star in the poorly understood final phases prior to core collapse \citep{Smith2014,CF17,PB17}.

Proper motion analysis is the most robust method for calculating the center of expansion (CoE) and explosion age of a SNR. Only a handful of known O-rich SNRs are sufficiently resolved to measure proper motion of high velocity ejecta from multi-epoch observations. This small list includes Cassiopeia A \citep{Kamper1976,Thorstensen2001,Fesen2006,Hammell2008}, Puppis A \citep{Winkler1985}, G292+1.8 \citep{Murdin1979,Winkler2009}, and 1E 0102.2-7219 (E0102) \citep{Finkelstein2006}, which is the focus of this paper. 

E0102 was discovered by the {\sl Einstein Observatory} during a survey of the Small Magellanic Cloud (SMC) \citep{SM1981}, and is approximately 62 kpc away \citep{Graczyk2014,Scowcroft2016}. E0102 was classified as an O-rich SNR \citep{Dopita1981} owing to its strong [O~III] $\lambda\lambda$4959, 5007 emission lines. Emission from other elements including Ar, Ne, C, Cl, Si, S, and Mg has also been identified \citep{Blair2000,Rasmussen2001,Seitenzahl2018,Alan2019}, with Ne and O being the most abundant \citep{Blair2000}. Localized hydrogen emission has been found in some knots \citep{Seitenzahl2018}, which is potentially consistent with a progenitor star partially stripped of its hydrogen envelope and a Type IIb supernova classification \citep{Filippenko97,Gal-Yam17,Sravan19}. The zero-age main-sequence mass estimates of E0102's progenitor ranges from 25-50 $M_{\odot}$ \citep{Blair2000,Flanagan2004,Finkelstein2006,Alan2019}.

The original estimate of E0102's explosion age was $\approx$1000 yr using a velocity map of [O III] $\lambda\lambda$4959, 5007 emission \citep{Tuhoy1983}. \cite{Hughes2000} calculated the percentage expansion of E0102 using three epochs of X-ray observations spanning 20 yr obtained with the \textit{Einstein}, \textit{ROSAT}, and \textit{Chandra X-ray Observatory}, and estimated an explosion age of $1000_{-200}^{+340}$ yr, consistent with \cite{Tuhoy1983}. However, a much older age of $\sim 2100$ yr was calculated using optical Fabry Perot imaging of oxygen-rich ejecta and fitting the velocity distribution with an ellipse \citep{Eriksen2001}.  \cite{Alan2019} used archival \textit{Chandra} data to estimate an explosion energy of $1.8 \times 10^{51}$ ergs and a Sedov Age of $\approx$3500 yr based on a forward shock velocity of 710 $\text{km s}^{-1}$, well above previous explosion age estimates. However, \cite{Xi2019}, also using archival \textit{Chandra} data, measured a forward shock velocity of $(1.61 \pm 0.37)\times10^{3}$ $\text{km s}^{-1}$  and estimated explosion ages of $\approx1700$\,yr or $\approx2600$\,yr depending on whether a constant or power law circumstellar density model is used.

\begin{figure*}[!thb]
\centering
\includegraphics[width=0.8\textwidth, angle=0]{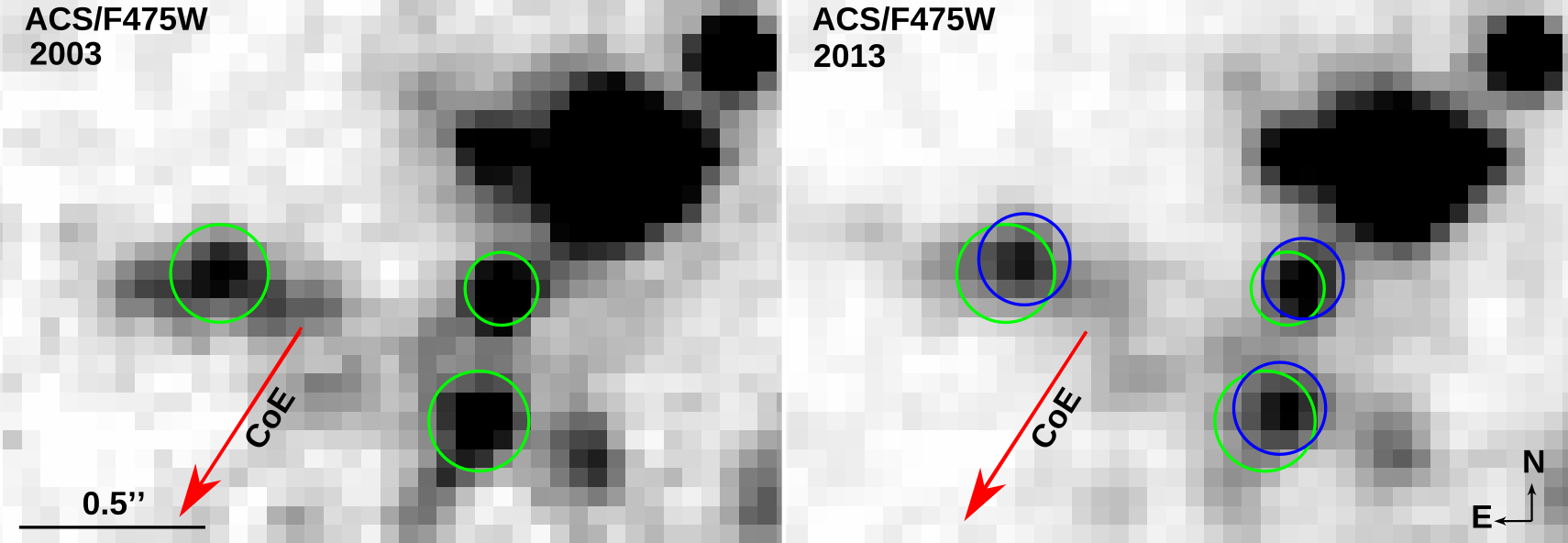}

\caption{An example of the expanding ejecta knots in 2003 (left) and 2013 (right). The 2003 knot centroids are shown as green circles while the 2013 centroids are shown as blue circles.}
\label{fig:Tracking}
\end{figure*}

\begin{deluxetable*}{lcccccccc}[!thb]
\label{tab:images}
\tablecaption{Observation information for the HST imaging of E0102}
\tablehead{
\colhead{PI} & \colhead{Date} & \colhead{Exp.\ Time} & \colhead{Instrument} & \colhead{Filter} & \colhead{$\lambda_{\rm center}$} & \colhead{Bandwidth} & \colhead{Velocity Range} & \colhead{Pixel Scale}\\
 & & (s) & & & (\r{A}) & (\r{A}) & ($\text{km s}^{-1}$) & $^{\prime\prime}$ pixel$^{-1}$\
}
\startdata
Morse & 07/04/1995 & 7200 & WFPC2/PC & F502N & 5012 & 27 & $\approx -$1000 to +1500 & 0.0455\\
Green & 10/15/2003 & 1520  & ACS/WFC & F475W & 4760 & 1458 & Full velocity range & 0.049\\
Madore & 04/10/2013 & 2044 & ACS/WFC & F475W & 4760 & 1458 & Full velocity range & 0.049\\
Milisavljevic & 05/12/2014 & 2753 & WFC3/UVIS & F502N & 5013 & 48 & $\approx -$2000 to +2500 & 0.040\\
Milisavljevic* & 05/12/2014 & 2665 & WFC3/UVIS & FQ492N & 4933 & 114 & Less than $-2000$ & 0.040\\
Milisavljevic* & 05/12/2014 & 2665 & WFC3/UVIS & FQ508N & 5091 & 131 & Greater than +$2500$ & 0.040\\
\enddata
\tablecomments{* denotes images not used in proper motion analysis}
\end{deluxetable*}

Among the most direct methods to estimate an explosion age is measuring proper motions of optically emitting dense knots of gas. \cite{Finkelstein2006} estimated the explosion age of E0102 to be 2054$\pm$584 yr from proper motion measurements of optically emitting ejecta observed in two \textit{Hubble Space Telescope} (\textit{HST}) images: a 1995 image using the Wide Field Planetary Camera 2 (WFPC2) and a 2003 image using the Advanced Camera for Surveys (ACS). \cite{Finkelstein2006} measured the proper motions of 12  regions and determined the CoE of the remnant to be $\alpha$=1$^{h}$04$^{m}$02.05$^{s}$ and $\delta$=-72$^{\circ}$01$^{\prime}$54.9$^{\prime\prime}$ (J2000) with a 1-$\sigma$ uncertainty of 3.4$^{\prime\prime}$ (henceforth Finkelstein CoE). This CoE is 2.4$^{\prime\prime}$ north and slightly east of a geometric center measured by fitting an ellipse to the X-ray bright shell \citep{Finkelstein2006}. 

Renewed interest in the precise location of the CoE of E0102 has been motivated by \cite{Vogt2018}, who report an X-ray source as a possible central compact object (CCO) of E0102 formed in the original supernova explosion. Using the Multi Unit Spectroscopic Explorer (MUSE) on the Very Large Telescope (VLT), they  discovered a ring of low-ionization Ne emission surrounding the X-ray source and concluded that the ring is being energized by the candidate CCO. The offset between the X-ray source and Finkelstein CoE implies a scenario where the CCO experienced a ``kick'' during the explosion with a transverse velocity of ~850 \kms\ \citep{Vogt2018}. The true nature of the X-ray source is unresolved. \cite{Rutkowski2010} had inspected this X-ray source using archival Chandra X-ray images to search for candidate CCOs, but did not find it to be credible. On the other hand, \cite{Hebbar2019} performed X-ray spectral analysis on the source and found that it could be a neutron star powered by strong magnetic fields ($B=10^{12}$ G). \cite{Long2020} suggest that the compact feature is not a point source and is a knot of ejecta (see Section \ref{sec:CCO} for more details). 

This paper improves over previous estimates of the CoE and the explosion age of E0102 by utilizing all available high resolution images obtained with {\sl HST} and a larger sample of proper motion measurements.  Section 2 discusses the images that were investigated, how they were measured for proper motion, and which epochs provided the most robust results. Section 3 describes our calculation of the CoE and explosion age and Section 4 discusses the implications of the measurements.

\section{Observations and Proper Motion Measurements} \label{sec:obs}

\begin{figure*}[!thb]
\centering
\includegraphics[width=0.95\textwidth, angle=0]{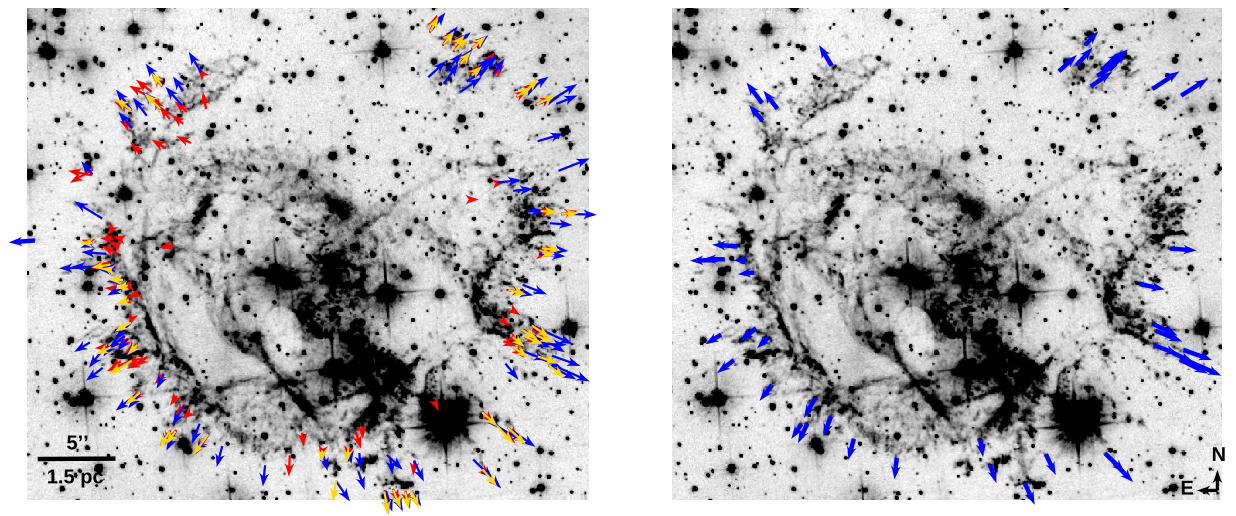}

\caption{Left: Vectors represent the measured shifts (multiplied by a factor of 20) of the two baselines and the multi-epoch set. Blue shows the 2003-2013 baseline, red shows the 1995-2014 baseline, and gold shows the multi-epoch set. Right: Vectors showing the selected knots used in CoE calculations.}

\label{fig:Center}
\end{figure*}

\begin{figure}[!htb]
\centering
\includegraphics[width=1\linewidth]{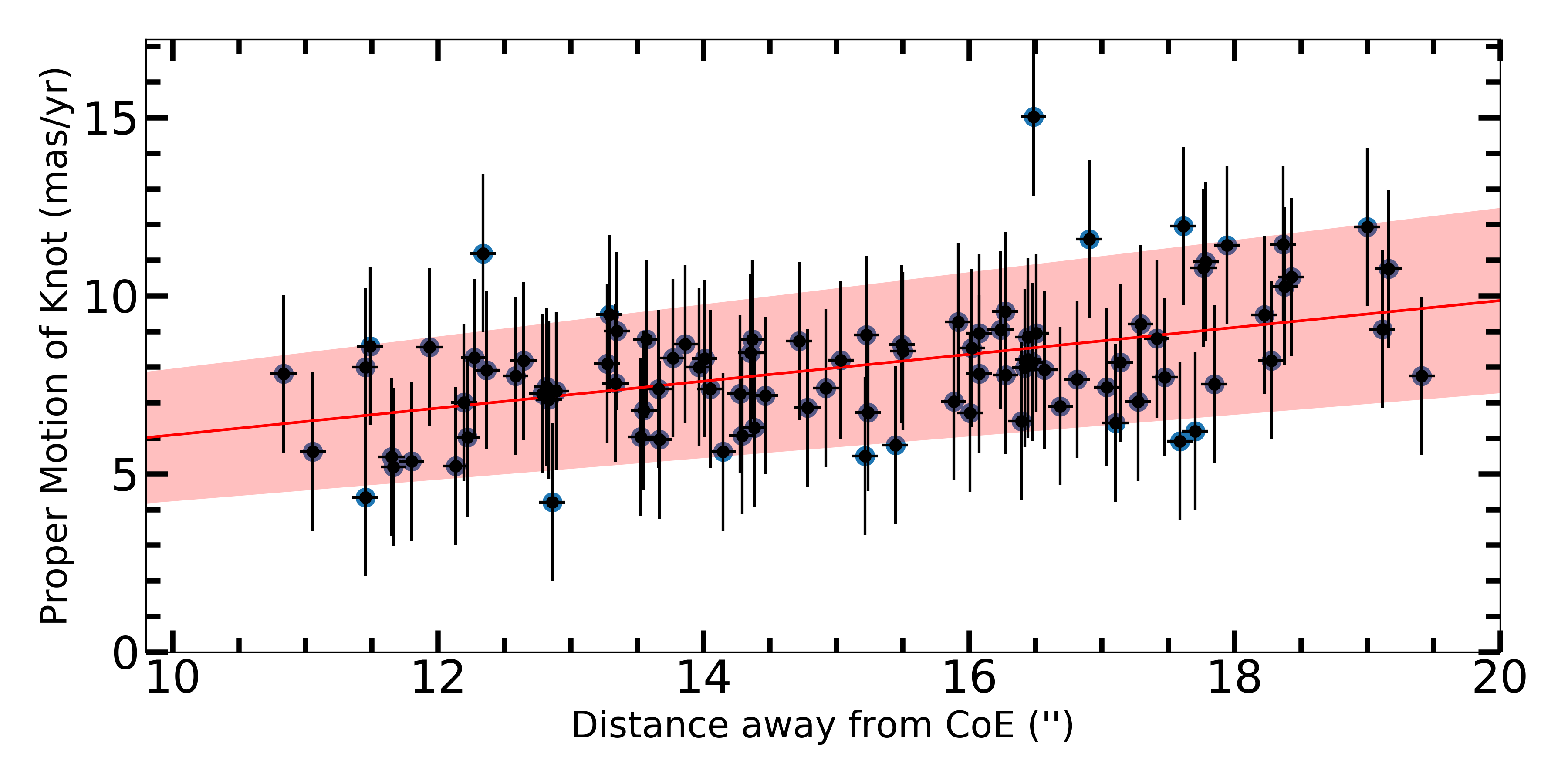} 

\caption{The absolute proper motion vs radial distance of the 2003-2013 baseline. The trend should follow a straight line in order to match a free expansion model.}

\label{fig:V_Radial}
\end{figure}

We examined four epochs of archival {\sl HST}  images of E0102, which were retrieved from the Mikulski Archive for Space Telescopes (MAST) and processed using \textit{Astrodrizzle}. The 1995 image was obtained with WFPC2, the 2014 image was obtained with the Wide Field Camera 3 (WFC3), and the 2003 and 2013 images were both obtained with ACS. The details of the images including the PIs, date of observations, the filters, their associated bandwidths, and pixel scales can be found in Table \ref{tab:images}. The F502N and F475W image filters are sensitive to emission from [O III] $\lambda\lambda$4959,5007. After processing, the image scale for all images is approximately 0.05$^{\prime\prime}$ pixel$^{-1}$. All images were cropped to fit a common 45$^{\prime\prime}$ $\times$ 45$^{\prime\prime}$ field of view. 

The images were aligned using the \texttt{geomap} and \texttt{geotran} tasks in IRAF\footnote{IRAF is distributed by the National Optical Astronomy Observatory, which is operated by the AURA, Inc., under cooperative agreement with the National Science Foundation. The Space Telescope Science Data Analysis System (STSDAS) is distributed by STScI.}. 
The \textit{geomap} command creates an image transformation database using anchor stars of two images, and \textit{geotran} applies the transformation. The anchors for the alignment can be found in Figure \ref{fig:geomapref}. 
Anchors were carefully chosen among stars with low residuals when \textit{geomap} was applied, excluding stars with high proper motions. Our anchor stars were among those used in \cite{Finkelstein2006}. 
Once the images were aligned, an accurate World Coordinate System was applied using a locally compiled version of the  \texttt{Astrometry.net\footnote{Astrometry is distributed as open source under the GNU General Public License and was developed on Linux}} \citep{Lang2010} code, which is accurate to $\approx0.2^{\prime\prime}$. 

\begin{figure*}[!th]
\centering
\includegraphics[width=0.7\textwidth, angle=0]{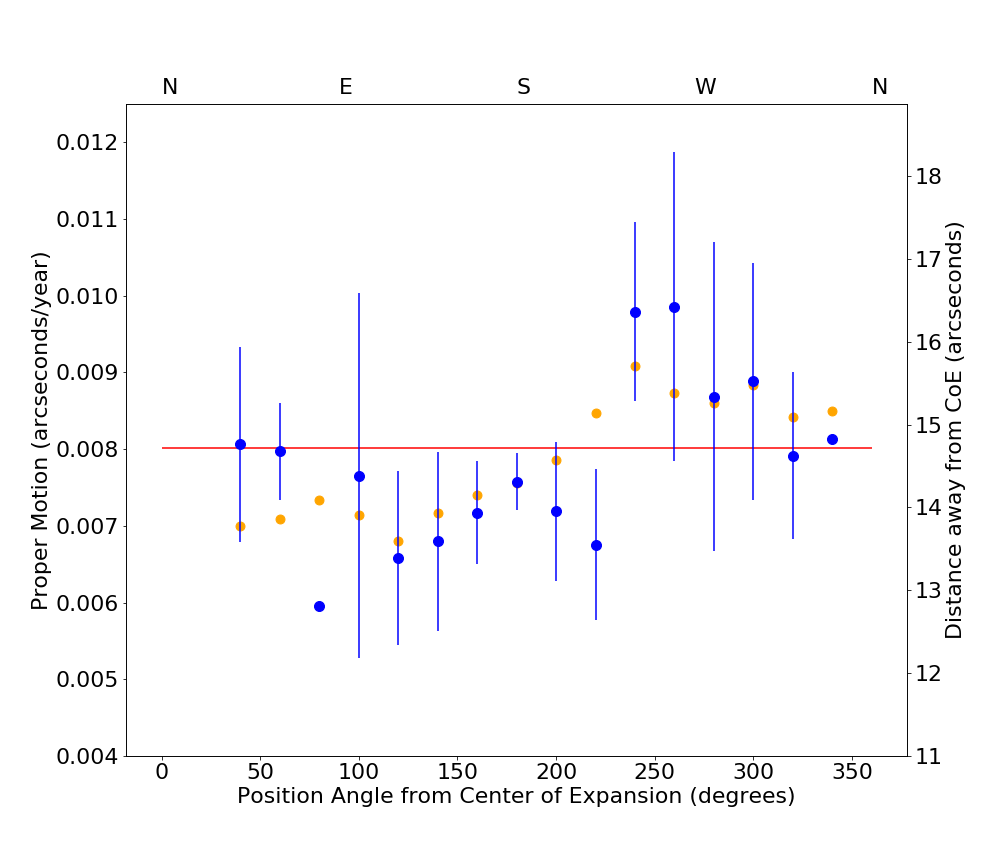}

\caption{Proper motion as a function of position angle of the 2003-2013 baseline knots. The blue points represent the average proper motion of the knots in a 20$^{\circ}$ slice, with the vertical blue lines showing the spread of the proper motion in the slice. The red horizontal line represents the average proper motion across the whole remnant. The orange points represent the average distance away from the CoE, shown on the right y-axis, and the associated proper motion assuming ballistic motion using values from Figure \ref{fig:V_Radial}. The position angle is from due North and sweeps counter-clockwise.}

\label{fig:V_R_T}
\end{figure*}

The individual knots were measured using two baselines, 2003-2013 and 1995-2014.  The 1995-2014 epochs (1995 WFPC2/F502N and 2014 WFC3/F502N) provide the longest baseline, whereas the 2003-2013 epochs (2003 ACS/F475W and 2013 ACS/F475W) were obtained using the same filter and instrument, which optimized tracking of individual knots. Knots were chosen by how well they could be tracked visually and their proximity to the edge of the remnant (larger than 8$^{\prime\prime}$ away from the Finkelstein CoE). The shifts of the knots were calculated by blinking between the two baseline images and visually locating the centers of knots or other conspicuous features (see Figure \ref{fig:Tracking}). The centers were measured multiple times to estimate positional errors ($\approx5\%$ relative error as compared to shifts). We measured 96 knots for the 2003-2013 baseline and 92 knots for the 1995-2014 baseline. A third multi-epoch data set was measured using all baselines. Implementing a similar multi-epoch measurement procedure as \cite{Winkler2009}, we measured 51 knots that were discernible in all of the epochs. All proper motion measurements can be seen in Figure \ref{fig:Center}.

We find that proper motion measurements made from images obtained with the same instrument and filter configurations were much more reliable and accurate than those made from different configurations with longer baselines. Although the 1995 and 2014 epochs provided the largest baseline, in multiple cases there was ambiguity as to whether knots were moving or brightening in new regions due to the sensitivity differences of the instruments and/or differences in resolution with shifts on the order of $\approx$ 1 pixel ($\approx0.05^{\prime\prime}$). The difference in bandpass between the two filters can be found in the Appendix (see Figure~\ref{fig:Thr}). The proper motion measurements for the 2003-2013 baseline have an average error of $\sim$20\%, significantly lower than the average error of $\sim$90\% in the 1995-2014 baseline, and $\sim$70\% for the multi-epoch data set. Thus, the knots of the 2003-2013 baseline were tracked with the highest level of confidence, making this baseline the optimal choice for proper motion analysis. Figure \ref{fig:V_Radial} shows our 2003-2013 proper motions measurements as a function of distance away from a CoE.

\section{Center of Expansion and Age} \label{sec:proper}
\begin{figure*}[!htb]
\centering
\includegraphics[width=0.48\textwidth, angle=0]{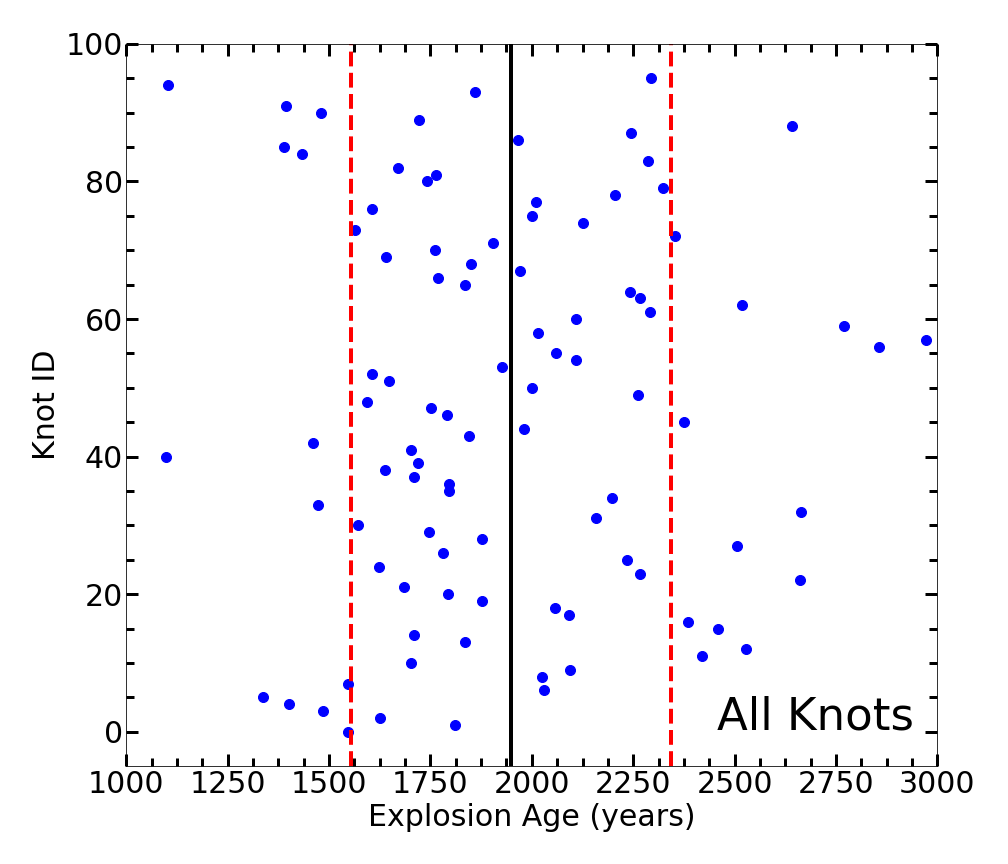}
\includegraphics[width=0.48\textwidth,angle=0]{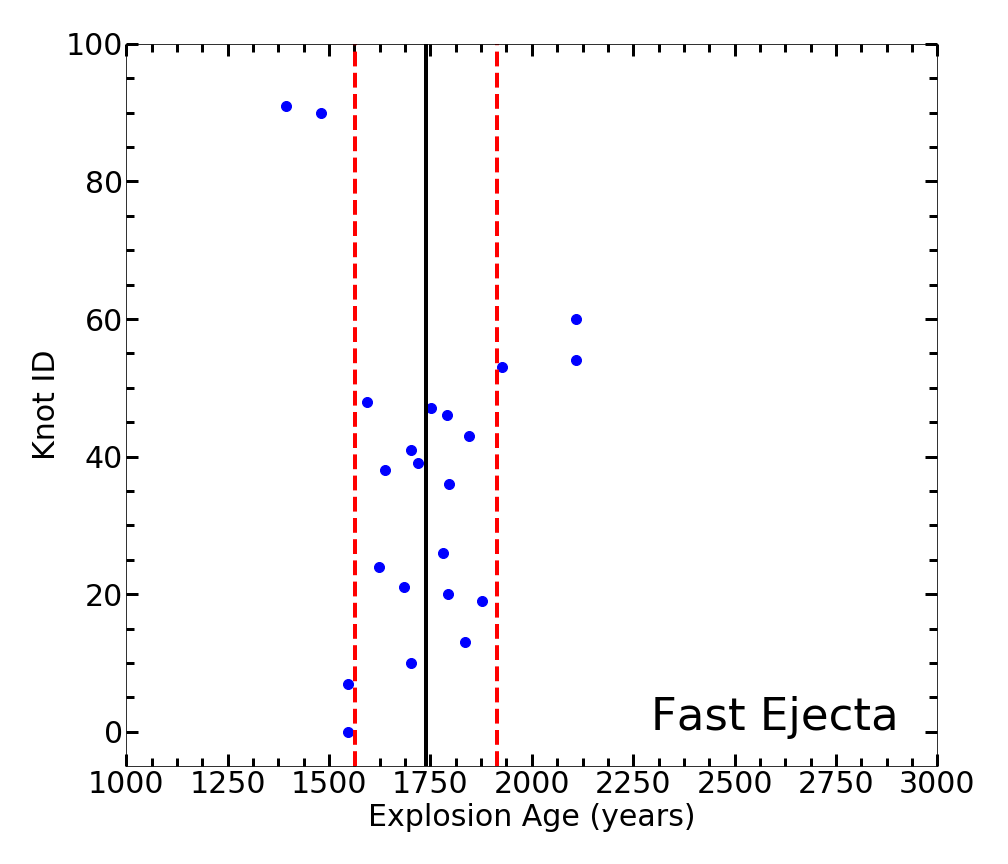}
\caption{A comparison of the explosions age measurements, assuming our CoE. The black line represents the average age of the data set, while the red dashed line is the 1-$\sigma$ uncertainty. Left: All the 2003-2013 baseline knots are used, resulting in an explosion age of $1948\pm395$ yr. Right: Only the fastest of the selected knots are used, resulting in an explosion age of 1738 $\pm$ 175  yr.}
\label{fig:Fesenyears}
\end{figure*}
\subsection{Proper motion asymmetry}
\label{sec:pmasymmetry}

Our approach of using many measurements of individual knots instead of measuring large regions allowed for the remnant to be reasonably well sampled along many position angles. This approach offers potential advantages over \citet{Finkelstein2006} who utilized a dozen large regions (each approximately $3-10$ square arcseconds in size) in order to compensate for the smaller baseline and differences in detector response and resolution. Consequently, \citet{Finkelstein2006} were only able to sample limited position angles.

Our measurements utilizing two epochs of ACS data were of sufficient resolution to identify that ejecta knots are not expanding uniformly, and that the rate of expansion changes with position around the remnant. In Figure \ref{fig:V_R_T}, the proper motion of knots as a function of the position angle, binned into 20$^{\circ}$ slices, is shown. The observed proper motion is compared to the expected proper motion when applying the 2003-2013 linear fit to the average distance away from the CoE of each slice (see additional details in Section \ref{sec:obs}). 
There is a clear division in observed versus expected proper motions between knots in the eastern versus western sides of the remnant. Between position angles $\approx$90-230$^{\circ}$, ejecta knots exhibit below average proper motion. A relationship between proper motion and position angle location of ejecta knots is an important consideration for techniques that assume ballistic motion and uniform expansion to determine the CoE and expansion age. Possible explanations for this non-uniform expansion are discussed in Section~\ref{sec:discussion}.

With this level of asymmetry present, knot selection becomes vitally important.
We carefully narrowed down the original 96 knots of the 2003-2013 baseline to 45 knots on the basis of tracking confidence, uniform shape between epochs, and trajectories that are within 20 degrees of the position angle from the Finkelstein CoE. These selected knots have been used to calculate the CoE and explosion age (right panel of Figure \ref{fig:Center}).

\begin{figure*}[!htb]
\centering
\includegraphics[width=0.7\linewidth, angle=0]{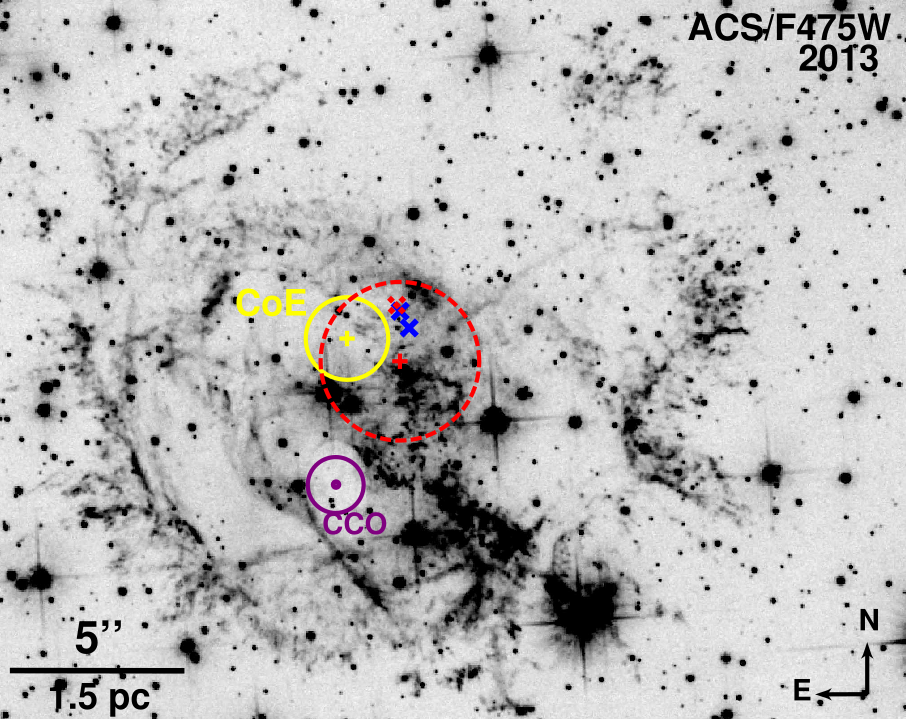}
\caption{Finkelstein CoE (red, dashed) and our result (yellow) with 1-$\sigma$ uncertainty circles. Also shown are the two X-ray geometric centers found by \cite{Xi2019} by matching ellipses to the forward and reverse shock (blue ``X"s), the geometric center from \cite{Finkelstein2006} (red ``X"), and \cite{Vogt2018}'s proposed CCO (purple) with its absolute uncertainty.}
\label{fig:Center_of_Expansion}
\end{figure*}

\begin{deluxetable*}{lccccc}[!th]
\tabletypesize{\small}
\label{table:CoE}
\tablecaption{Center of Expansion and Explosion Age Estimates}
\tablehead{
\colhead{Reference} & \colhead{CoE $\alpha$ (J2000)} & \colhead{CoE $\delta$ (J2000)} & \colhead{1-$\sigma$ uncertainty ($^{\prime\prime}$)}  & \colhead{Explosion Age (yr)}
}
\startdata
This Paper & 1$^{h}$04$^{m}$02.48$^{s}$ & -72$^{\circ}$01$^{\prime}$53.92$^{\prime\prime}$ & 1.77 & 1738$\pm$175 \\
\cite{Finkelstein2006} & 1$^{h}$04$^{m}$02.05$^{s}$  & -72$^{\circ}$01$^{\prime}$54.9$^{\prime\prime}$ & 3.4 & 2054$\pm$584\\
X-ray Center \citep{Finkelstein2006}& 1$^{h}$04$^{m}$02.08$^{s}$ & -72$^{\circ}$01$^{\prime}$52.5$^{\prime\prime}$ & \nodata & \nodata \\
Forward Shock Center \citep{Xi2019}& 1$^{h}$04$^{m}$01.964$^{s}$ & -72$^{\circ}$01$^{\prime}$53.47$^{\prime\prime}$ & \nodata & \nodata \\
Reverse Shock Center \citep{Xi2019} & 1$^{h}$04$^{m}$02.048$^{s}$ & -72$^{\circ}$01$^{\prime}$52.75$^{\prime\prime}$ & \nodata & \nodata \\
\cite{Tuhoy1983} & \nodata & \nodata & \nodata & 1000 \\
\cite{Hughes2000} & \nodata & \nodata & \nodata & $1000_{-200}^{+340}$\\
\cite{Eriksen2001} & \nodata & \nodata & \nodata & 2100 \\
\enddata
\end{deluxetable*}

\subsection{Center of Expansion}
\label{sec:distance}

Our approach to determine the CoE of E0102 uses selected proper motion measurements of the 2003-2013 baseline in combination with a likelihood function. This method is similar to that used by \cite{Thorstensen2001} to determine the CoE of Cassiopeia A. We favor this method because it only depends on the direction of the knots, and is not sensitive to deceleration or non-uniform expansion over time. Other methods \citep[e.g.][]{Winkler2009} were considered. However, these methods assume uniform expansion, which cannot be assumed with E0102.

The likelihood function used is:
\begin{equation*}
    \lambda(X,Y)=\Pi_{i}\frac{1}{2\sigma_{i0}}exp(-d^{2}_{i\perp}/2\sigma_{i0}^{2}),
\end{equation*}
where (X,Y) is an arbitrary center of expansion and $d_{i \perp}$ is the perpendicular distance between (X,Y) and the knot's line of position \citep{Thorstensen2001}. 
We defined $\sigma_{i0}$ to be the 
uncertainty associated with the point common to the knot's extended line of position and $d_{i \perp}$. This differs from the original likelihood function from \cite{Thorstensen2001}, where $\sigma_{i0}$ was defined to represent the positional uncertainty of the knot near the time of explosion. We modified $\sigma_{i0}$ because of our comparatively poor constraint on the time of explosion.
The (X,Y) that maximizes this function gives the CoE. This is repeated for 100,000 artificial data sets generated from position and direction distributions of individual knots (see Section \ref{sec:obs}). 
Using this method and proper motion measurements from the selected 45 knots 
yields a CoE of
$\alpha$=1$^{h}$04$^{m}$02.48$^{s}$
and $\delta$=-72$^{\circ}$01$^{\prime}$53.92$^{\prime\prime}$ (J2000) with 1-$\sigma$ uncertainty of 1.77$^{\prime\prime}$.

To confirm the purity of our selected 45 knots used in our estimate of the CoE, we ran an additional 100 calculations of the CoE using 45 randomly selected knots from all 96 measurements for each iteration. The resulting CoEs are within 1-$\sigma$ in radius around our favored CoE, but the majority are $0.5^{\prime\prime}$ away and collectively are associated with a larger average error (approximately $2^{\prime\prime}$). Thus, narrowing down knot measurements from the original 96 to 45 introduces a non-negligible improvement in our estimate of the CoE.

\subsection{Explosion Age}
\label{sec:fastejecta}
Our method to determine the explosion age uses only knots with the highest proper motions to calculate an explosion age. A similar approach was used by \citet{Fesen2006} to improve upon the explosion age of Cassiopeia~A, first made in \citet{Thorstensen2001}. Proper motions of ejecta in E0102 are not uniform around the remnant, which suggests that some regions have been decelerated and more strongly influenced by interaction with surrounding material.

By assuming that knots with the fastest proper motions are least decelerated and therefore truer representations of the initial ejection velocity of the knots, a more accurate explosion age can be inferred. 22 of the 45 selected knots with proper motions greater than the average (0.008$^{\prime\prime}$ yr$^{-1}$) were examined. These knots represent the highest proper motions within E0102. The explosion age was calculated by dividing a knot's distance away from the CoE by the proper motion of the knot. Adopting the CoE derived from the likelihood method (Section \ref{sec:distance}), this resulted in an explosion age of 1738 $\pm$ 175  yr. 
The difference in choosing all of the knots versus the fastest can be found in Figure \ref{fig:Fesenyears}. If all 96 knots from the 2003-2013 baseline are included in the calculation, the result is $1948\pm395$ yr, similar to \cite{Finkelstein2006}. Notably, if we further restrict our selection of knots to only those having the fastest proper motion and that are furthest out from the CoE (corresponding to knots between position angles 200-280$^{\circ}$), we still retrieve our favored explosion age ($\approx1740$ yr). Our fast ejecta explosion age is  consistent with the age found in \cite{Xi2019} using a constant density model for the ambient medium (see Section \ref{sec:asymmetry} for our interpretation).

\section{Discussion} \label{sec:discussion}

\subsection{Knot Results}
\label{subsec:results}

In Table \ref{table:CoE} we list the results of our CoE and explosion age estimates along with estimates made by previous studies.  Figure \ref{fig:Center_of_Expansion} shows the coordinates of our CoE, the Finkelstein CoE, and the proposed CCO \citep{Vogt2018}, with their associated uncertainties. Our CoE is approximately 2.0$^{\prime\prime}$ east and 1.0$^{\prime\prime}$ north of the Finkelstein CoE, and the two estimates are  consistent within uncertainties.

Notably, our CoE and 1-$\sigma$ error estimate lies roughly 2.3$^{\prime\prime}$ away from estimates of E0102's geometric X-ray center and reverse shock's geometric center, and $\approx$2.4$^{\prime\prime}$ away from the forward shock's geometric center \citep{Finkelstein2006,Xi2019}. An offset between the CoE and the geometric X-ray center is not unique to E0102, and is observed in Cassiopeia~A, G292.0+1.8, and Puppis A  \citep[references therein]{Katsuda2018}. One explanation for this discrepancy is age and the associated prolonged interaction with CSM  \citep{Katsuda2018}. This likely applies to E0102 as well, given the asymmetry in the expansion we observe (Figure~\ref{fig:V_R_T}).

\subsection{Ablation trails}

\begin{figure}[ht]
\centering
\includegraphics[width=\linewidth, angle=0]{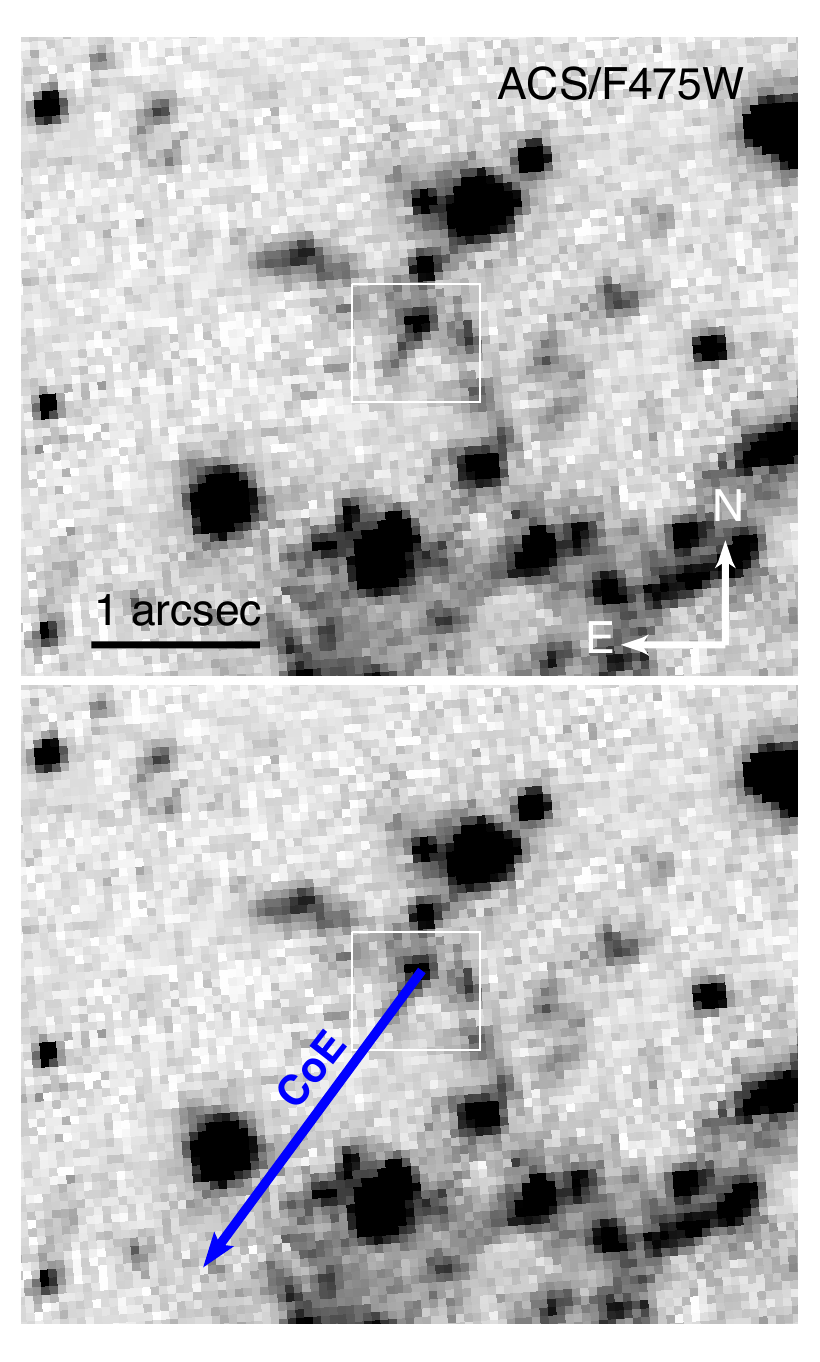}

\caption{Example knot from the 2003-2013 baseline exhibiting an ablation trail tracing back to our estimated CoE.}

\label{fig:trail}
\end{figure}

An additional check for our calculated CoE is to inspect alignment between the ablation trails of ejecta knots, if visible, with radial vectors extending from the CoE to the ejecta knot's location. \citet{Fesen2011} demonstrated how the gradual dissolution of high-velocity ejecta caused by passage through CSM/ISM in Cassiopeia A can leave trailing emission that traces back to the CoE. Figure \ref{fig:trail} shows an example knot in E0102 where the ablation trail is visible and has a path that traces back towards our estimated CoE. There are fewer examples of ejecta knot ablation in E0102 as compared to Cassiopeia A, which is likely a consequence of E0102's advanced age and greater distance.  Ejecta knots observed in Cassiopeia A are $0.2-1.0^{\prime\prime}$ ($1-5 \times 10^{16}\, \rm cm$) in size, and thus at the distance of E0102 we are only sensitive to the largest and most brightly emitting knots.  

\subsection{Expansion asymmetry}
\label{sec:asymmetry}
\begin{figure*}[!htb]
\centering
\includegraphics[width=0.47\textwidth, angle=0]{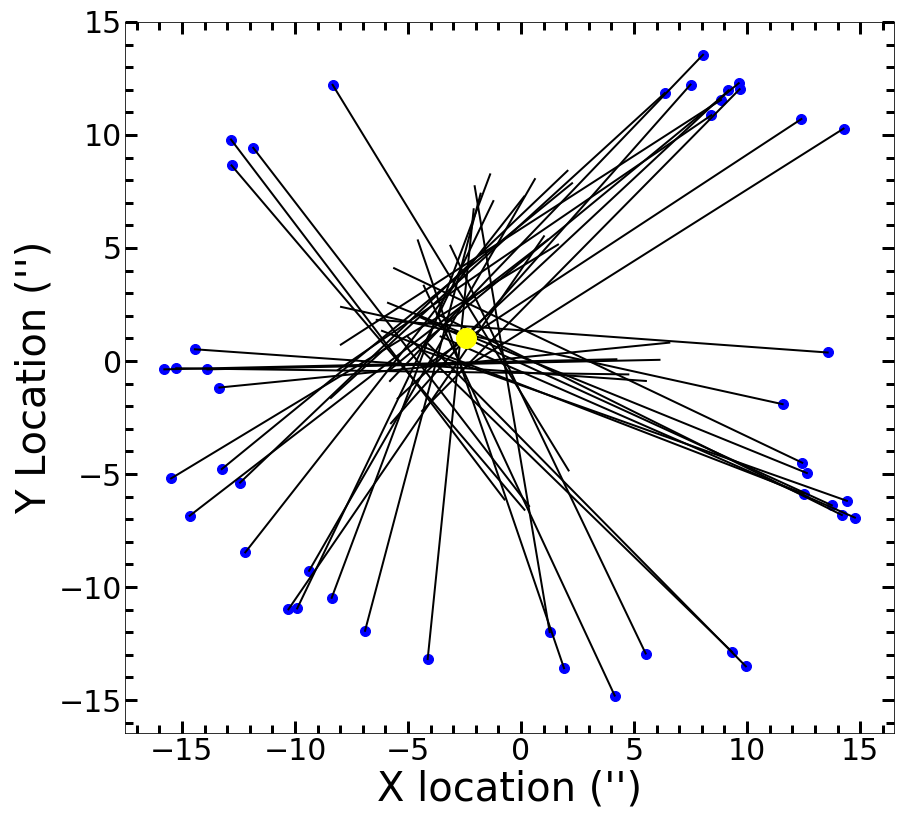}
\includegraphics[width=0.47\textwidth, angle=0]{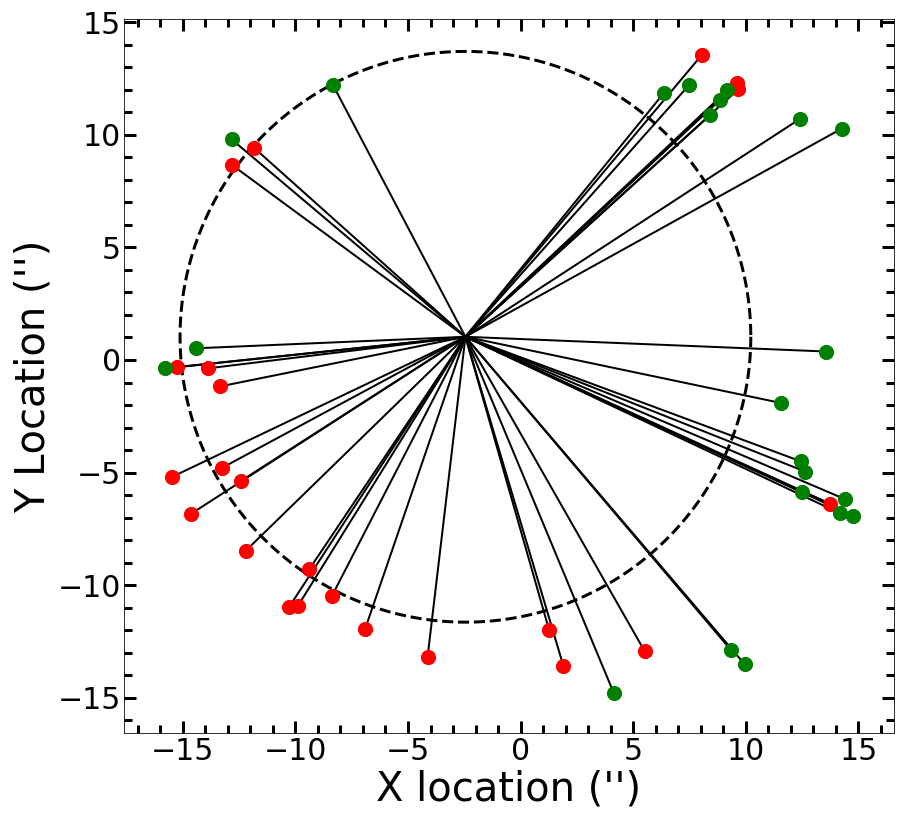}
\caption{Left: The proper motions of the 45 selected knots traced back 20$^{\prime\prime}$ ($\approx$2500 yr assuming the average proper motion of 0.008$^{\prime\prime}$\,yr$^{-1}$). The CoE is shown in yellow. Right: The trajectories of the selected knots if forced to originate from the CoE. The green points indicate the knots used for the Fast Ejecta explosion age estimate, while the red points were not.  The dashed black line shows the expansion if constrained to the average global proper motion of 0.008$^{\prime\prime}$\,yr$^{-1}$.}

\label{fig:Fesen}
\end{figure*}

Our proper motion measurements of E0102 using two epochs of {\sl HST}/ACS images were of sufficient angular and temporal resolution to determine that ejecta knot expansion is not homologous. With high confidence we find asymmetry in the proper motions as a function of position angle (see Figure~\ref{fig:V_R_T}). This phenomenon is highlighted in Figure \ref{fig:Fesen}. The left panel of Figure \ref{fig:Fesen} shows knot trajectories tracing back 2500 yr as compared to our CoE. The right panel shows the high proper motion knots used for both our age estimate and our CoE calculation (shown in green), the knots used for the CoE calculation and not the explosion age estimate (red), and the shape of the remnant if expanding uniformly at the mean proper motion ($0.008^{\prime\prime}$ yr$^{-1}$; dashed circle). 
While average uniform expansion fits with the eastern limb, it does a poor job of fitting the western portion of the remnant. The discrepancy provides strong evidence for non-uniform expansion.  The non-uniform expansion may be due to either asymmetry in the original supernova explosion or interaction between the original supernova blast wave and an inhomogeneous CSM/ISM. 

Explosion asymmetry has been suggested for E0102. \cite{Finkelstein2006} noticed a possible jet structure running SW-NE and a possible density gradient within the CSM, caused by mass loss of the progenitor. They proposed a Wolf-Rayet (WR) progenitor star after examining the surrounding environment, especially the possible association with N76A, a nearby hydrogen cloud with characteristics similar to a WR bubble. However, \cite{Vogt2010} found that the association with N76A is unlikely due to increasing ISM density along the path to N76A. A perceived preferred axis in their 3D reconstruction and the ISM density gradient being perpendicular to this axis led \cite{Vogt2010} to favor explosion asymmetry.

However, our measurements highlight a conspicuous correlation between knots with slower-than-average proper motion and regions of increased supernova--CSM interaction. In particular, X-ray studies have shown that the southeast portions of the remnant where we measure the slowest proper motions is also a region of increased CSM density \citep{Sasaki2006}. Likewise, the forward shock has a greater extent in the  southwest \citep{Xi2019}, where we observe some of the fastest proper motions. Thus, ejecta knot expansion asymmetry is most likely due to interaction between the original supernova and an inhomogeneous CSM/ISM that has decelerated ejecta in localized regions.

\begin{figure*}[!htp]
\centering
\includegraphics[width=0.90\textwidth, angle=0]{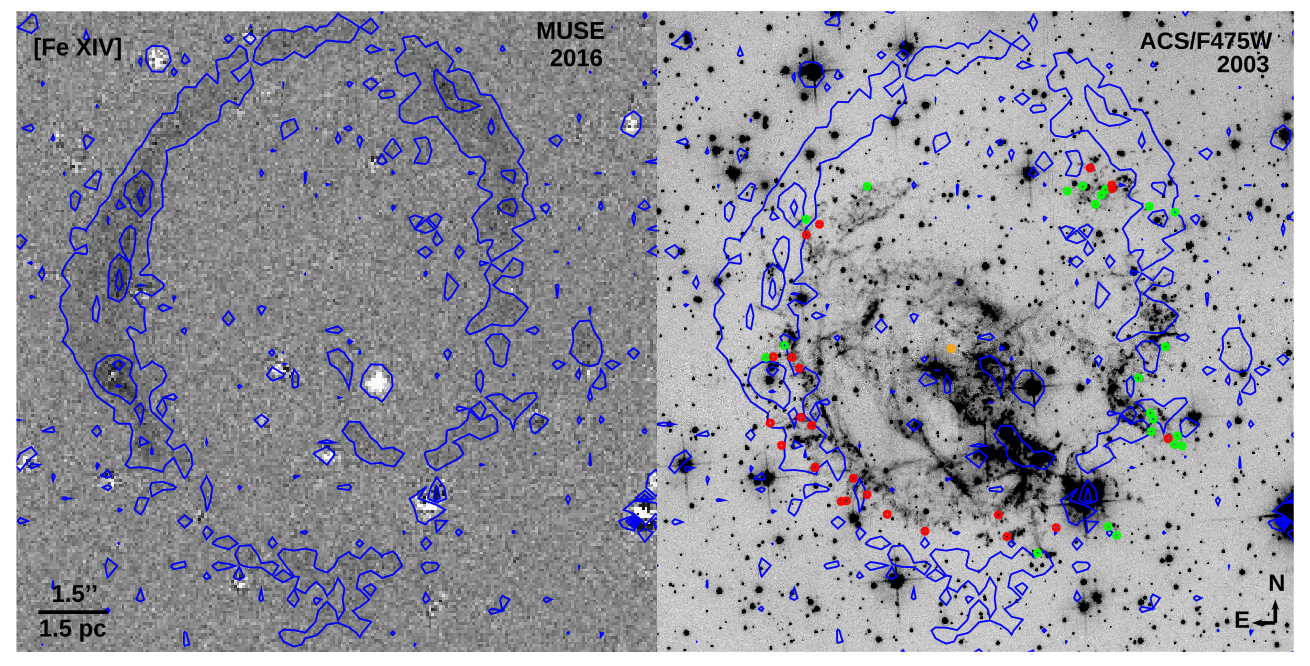}

\caption{Left: Image and associated contour plot of the integrated [Fe XIV] emission from Figure 2 of \cite{Vogt2017}. Right: 2003 ACS/F475W image comparing the Fe emission found in \cite{Vogt2017} (contours show in blue) and the proper motion of the 2003-2013 baseline knots. The green points are the knots used in calculating the Fast Ejecta explosion age estimate, while the red points identify knots that were not used in the explosion age estimate. The orange point shows our CoE. }
\label{fig:Fe_ejecta}
\end{figure*}

The inhomogeneous environment would have been sculpted by the mass loss of the massive star progenitor.   \citet{Finkelstein2006} suspected that the E0102 SNR was expanding into an asymmetric bubble swept out by the strong winds of a WR progenitor star. This interpretation was supported by spectroscopic observations made with MUSE sensitive to [Fe~XIV] $\lambda5303$ and [Fe~XI] $\lambda7892$ by \cite{Vogt2017} that appear to trace this remnant bubble (see Figure~\ref{fig:Fe_ejecta}).  These emission lines, which map forward shock interaction with dense ISM, correlate with regions where O-rich ejecta have lower proper motion. This is particularly noticeable in the east where an extended filament of ejecta with lower-than-average proper motion overlaps with strong [Fe~XIV] emission. Likewise, knots with higher-than-average proper motion are disproportionately located in the southwest where there is the least amount of [Fe~XIV] emission. 

Taking this expansion asymmetry into consideration, we used the fastest and presumably least decelerated knots to calculate E0102's expansion age to be 1738 $\pm$ 175  yr. This is comparable with \cite{Eriksen2001} and \cite{Finkelstein2006} age estimates, but outside of the \cite{Hughes2000} estimate. Our age estimate should be strictly interpreted as an upper limit because we are unable to measure the amount of deceleration experienced by the ejecta knots used in our proper motion analysis. Additional epochs of proper motion measurements could potentially determine the deceleration experienced by the knots, giving a more precise age estimate.

Our explosion age estimate is also consistent with the age calculated by \cite{Xi2019} using a global constant density ambient medium, combining mass loss and ISM. However, the scenario favored by our data is more complex, such that the eastern expansion of ejecta has encountered ISM/CSM gas of higher density. \cite{Xi2019} also notes that with their models, a WR progenitor isotropically losing mass is unlikely. The mass loss could be caused by a single star progenitor through episodic eruptions  \citep[see][for discussion]{Smith2014}. Another channel of mass loss is via inefficient mass loss in binary interactions \citep{Ouchi2017,Sravan2020}. Recently, \citet{Seitenzahl2018} reported the detection of hydrogen spectral features within E0102, supporting a Type IIb progenitor. It has been found that the majority of progenitor stars of Type IIb SNe are partially stripped of their hydrogen-rich envelopes via binary interactions \citep{Claeys2011,Yoon2017,Sravan19}.

\subsection{Proposed CCO Results}
\label{sec:CCO}

Our CoE is $\approx$6.4$^{\prime\prime}$ away from the candidate CCO proposed by \cite{Vogt2018}. Assuming a distance of 62 kpc \citep{Graczyk2014,Scowcroft2016}, this would correspond to $1^{\prime\prime}\approx$ 0.3 pc. Using our  calculated CoE and explosion age, this translates to a kick velocity of $1070 \pm 380$ $\text{km s}^{-1}$ for the proposed CCO. This velocity is larger but still within uncertainties of the velocity  calculated by \cite{Vogt2018} using the Finkelstein CoE and age ($\approx850$ $\text{km s}^{-1}$).

The inferred kick velocity is large compared to kick velocities of other neutron stars. \cite{Hobbs2005} found that in a sample of 233 pulsars, most pulsars younger than 3 Myr have a mean velocity of just 400$\pm$265 $\text{km s}^{-1}$. Only two pulsars have velocities above 1000 $\text{km s}^{-1}$, both with questionable distance estimates. However, E0102's inferred kick velocity is comparable to two other young SNRs, 
Puppis A and N49, with plane of sky velocities of 763$\pm$73 $\text{km s}^{-1}$ \citep{Mayer2020} and 1100$\pm$50 $\text{km s}^{-1}$ \citep{Katsuda2018}, respectively.  It should be noted that the velocity for N49 was found using the X-ray center and not a CoE, which could lead to an overestimation (see Section \ref{subsec:results} for more details).

On the other hand, simulations have shown that kick velocities of over 1000 $\text{km s}^{-1}$  may be possible. \cite{Scheck2006} observed neutron stars with kick velocities exceeding 1000 $\text{km s}^{-1}$  in 2D simulations. In 3D models, \cite{Wongwathanarat2013} found that core collapse explosions of progenitor stars with ZAMS between 15-20$M_{\odot}$ could generate kicks with velocities of upward of 700 $\text{km s}^{-1}$. \cite{Janka2017} found that, for 2D and 3D simulations, a high level of explosion asymmetry (common in high mass progenitor systems) can cause higher kick velocities.

 \cite{Hebbar2019} and  \cite{Long2020} each conducted a careful analysis of Chandra  observations of the candidate CCO that included time-dependent responses for each of the archival observations, modeling of the background instead of subtracting it, and fitting the unbinned spectra to preserve the maximal spectral information.  \cite{Hebbar2019} found that a single blackbody model does not provide an acceptable fit to the data, but a blackbody+power-law or a neutron star carbon atmosphere model do provide acceptable fits. However, the  blackbody+power-law model fit has a neutral hydrogen column density that is  $\sim10\times$ higher than the accepted value to E0102 and the neutron star carbon atmosphere model has a value that is $\sim18\times$ higher. \cite{Long2020} also found that a single blackbody model is not consistent with the data and they fit the spectrum of the compact feature with a thermal, non-equilibrium ionization model, finding acceptable fits with neutral hydrogen column density values consistent with the accepted value.  They find two classes of thermal models can fit the spectra equally well, one with a temperature of $kT\sim0.79$~keV, an ionization timescale of $\sim3\times10^{11}\,\mathrm{cm}^{-3}\mathrm{s}$, and marginal evidence for enhanced abundances of O and Ne, and the other with a temperature of $kT\sim0.91$~keV, an ionization timescale of $\sim7\times10^{10}\,\mathrm{cm}^{-3}\mathrm{s}$, and abundances consistent with local interstellar medium values. The limited statistics in the spectrum of this faint feature prevents any further discrimination amongst these spectral models. 
 
\cite{Long2020} also conducted an analysis of the spatial distribution of the counts, and showed that the distribution is not consistent with that of an isolated point source. Though they could not rule out a point source embedded in a region of diffuse emission, its flux must be significantly lower than the values reported in \cite{Vogt2018} and \cite{Hebbar2019}. Based on the spectral and image analysis, \cite{Long2020} questioned the association of the X-ray source with a neutron star and suggest instead that it is likely to be a knot of O- and Ne-rich ejecta associated with the reverse shock. 

\section{Conclusion} \label{sec:conclusion}

We have estimated the CoE and expansion age of E0102 by measuring proper motions of O-rich ejecta observed in multi-epoch {\sl HST} images. We analyzed all [O~III]-sensitive images over a 19 year period, but found that the 2003-2013 ACS/F475W baseline from which 45 different knots could be confidently tracked produced the most robust results.
The high resolution of {\sl HST} made it possible to identify evidence of non-homologous expansion of the knots, which we conclude to be the result of interaction with an inhomogeneous CSM environment. 

We calculated the CoE using only the direction of the proper motion and a likelihood function, which yielded a CoE of $\alpha$=1$^{h}$04$^{m}$02.48$^{s}$ and $\delta$=-72$^{\circ}$01$^{\prime}$53.92$^{\prime\prime}$ (J2000) with the 1-$\sigma$ uncertainty being 1.77$^{\prime\prime}$. This CoE is  2.2$^{\prime\prime}$ away from, but consistent with, the CoE calculated by \cite{Finkelstein2006}. Using only the
fastest knots we calculated an explosion age of 1738 $\pm$ 175  yr. Our CoE is 6.4$^{\prime\prime}$  away from the candidate CCO proposed by \citet{Vogt2018}, implying a
transverse kick velocity of $\approx$1070 $\text{km s}^{-1}$. This is an unusually large velocity compared to an
average neutron star velocity of 400 $\text{km s}^{-1}$ \citep{Hobbs2005} and one of the highest
among other SNRs \citep{Katsuda2018}, although simulations have been able to achieve kick velocities greater than 1000 $\text{km s}^{-1}$ \citep[e.g.][]{Scheck2006}. Our results generally support the recent conclusions of \citet{Long2020} that the X-ray source identified as a CCO may well be a knot of ejecta that has been excited by the reverse shock. 

A new epoch of HST/ACS images would expand our understanding of E0102. Such images would enable multi-epoch analysis that would tighten uncertainties on proper motion, estimate potential deceleration, and further constrain the CoE and explosion age. This, in turn, would test our conclusion that non-homologous expansion of E0102's optical knots is caused by interaction of the original supernova blast wave with inhomogeneous CSM. 

\acknowledgements

D.~M.\ acknowledges NSF support from grants PHY-1914448 and AST-2037297. This research is based on observations made with the NASA/ESA Hubble Space Telescope obtained from the Space Telescope Science Institute, which is operated by the Association of Universities for Research in Astronomy, Inc., under NASA contract NAS 5-26555. These observations are associated with HST programs 6052, 12001, 12858, and 13378. Support for program \#13378 was provided by NASA through a grant from the Space Telescope Science Institute, which is operated by the Association of Universities for Research in Astronomy, Inc., under NASA contract NAS 5-26555. F.\ Vogt provided MUSE observations, processed with \texttt{brutifus} \citep{brutifus}, and helpful comments on an earlier draft of the manuscript.

\software{PYRAF \citep{PYRAFcite}, ds9 \citep{ds9cite}, astrometry.net \citep{Lang2010}, Astropy \citep{AstropyCiteA,AstropyCiteB}}

\bibliography{bib}{}
\bibliographystyle{aasjournal}

\hfill \break

\appendix
\setcounter{figure}{0}
\renewcommand{\thefigure}{A\arabic{figure}}

We include here plots showing the total system throughput of the various {\it HST} instrument + filter configurations used to image the O-rich ejecta of E0102.

\begin{figure*}[!ht]
\centering
\includegraphics[width=0.4\textwidth, angle=0]{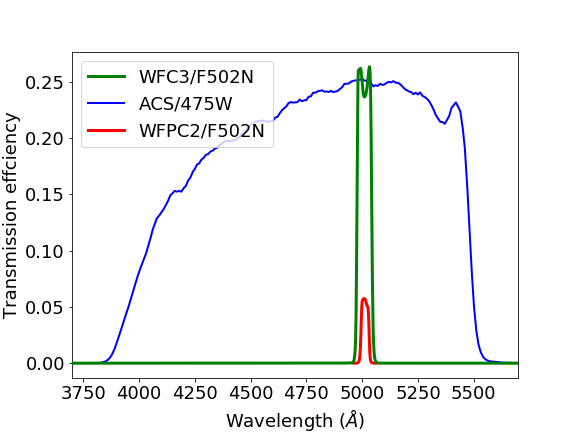}
\includegraphics[width=0.4\textwidth, angle=0]{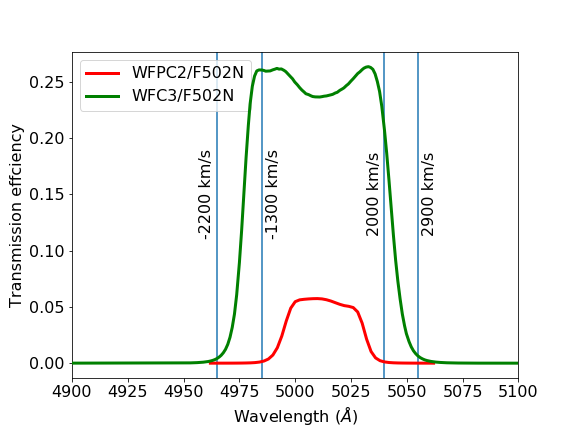}
\caption{Left: Total system throughput efficiency plots of {\sl HST} instrument + filter combinations used to image O-rich ejecta of E0102 that were analyzed in this paper. Right: Total system throughput efficiency plots enlarged around the two F502N filters of WFPC2 and WFC3. The sensitivity differences introduce uncertainty in tracking proper motion of knots between the 1995-2014 baseline.}
\label{fig:Thr}
\end{figure*}

\end{document}